\documentclass[12pt]{article}
\usepackage{amsmath}
\usepackage{amssymb}
\begin{document}
\vskip 2cm
\begin{center}
{\sf { \Large  A Specific $\mathcal{N }= 2$ Supersymmetric Quantum Mechanical Model: Supervariable Approach}}

\vskip 1.0cm

{\sf Aradhya Shukla}\\
{\it Indian Institute of Science Education and Research Kolkata,\\ Mohanpur-741246, India}\vskip 0.2cm
{\small {\sf {e-mails: as3756@iiserkol.ac.in;  ashukla038@gmail.com}}}

\end{center}

\vskip 2cm

\noindent
{\bf Abstract:} By exploiting the supersymmetric invariant restrictions on the chiral and anti-chiral supervariables, 
we derive the off-shell nilpotent symmetry transformations for a specific (0 + 1)-dimensional 
$\mathcal{N} = 2$ supersymmetric quantum mechanical model which is considered on a (1, 2)-dimensional 
supermanifold (parametrized by a bosonic variable $t$ and a pair of Grassmannian
variables ($\theta,\, \bar\theta$). We also provide the geometrical meaning to the symmetry transformations. 
Finally, we show that this specific $\mathcal {N} = 2$ SUSY quantum mechanical model is a model for  Hodge theory. 



\newpage

\section{Introduction}

Gauge theory is one of the  most important theories of modern physics because {\it three} out of {\it four} fundamental interactions of nature 
are governed by the gauge theory. The Becchi-Rouet-Stora-Tyutin (BRST) formalism is one of the systematic approaches to covariantly quantize any
$p$-form ($p =1, 2, 3...$) gauge theories, where the local gauge symmetry of a given theory is traded with the ``quantum'' gauge (i.e. (anti-)BRST)
symmetry transformations [1-4]. It is important to point out that, the (anti-)BRST  symmetries are nilpotent and absolutely anticommuting in nature.
One of the unique, elegant and geometrically rich methods to derive these (anti-)BRST transformations is the superfield formalism, where 
the horizontality condition (HC) plays an important role [5-12]. This HC is a useful tool to derive the BRST, as well as anti-BRST symmetry transformations 
for any (non-)Abelian $p$-form ($p =1, 2, 3...$) gauge theory, where no interaction between the gauge and matter fields are present.

For the derivation of a full set of (anti-)BRST symmetry transformations in the case of interacting gauge theories, a
powerful method known as augmented version of superfield formalism has been developed in a set of papers [13-16]. In augmented 
superfield formalism, some conditions named as gauge invariant restrictions (GIRs), in addition to the HC, have been imposed 
to obtain the off-shell nilpotent and absolutely anticommuting (anti-)BRST transformations. It is worthwhile to mention 
here that this technique has also been applied in case of some $\mathcal {N} = 2$ supersymmetric (SUSY) quantum mechanical (QM) models to derive 
the off-shell nilpotent SUSY symmetry transformations [17-20]. These SUSY transformations have been derived by 
using the supersymmetric invariant restrictions (SUSYIRs) and it has been observed that the SUSYIRs are the 
generalizations of the GIRs, in case of $\mathcal {N} = 2$ SUSY QM theory.

The aim of the present investigation is to explore and apply the augmented version of HC to a  {\it new}
$\mathcal {N} = 2$ SUSY QM model which is different from the earlier models present in the literature. In our present endeavor, we derive 
the off-shell nilpotent SUSY transformations for a specific $\mathcal {N} = 2$ SUSY QM model by exploiting the 
potential and power of the SUSYIRs. The additional reason behind our present investigation is 
to take one more step forward in the direction of the confirmation of SUSYIRs (i.e., generalization of augmented superfield formalism) 
as a powerful technique for the derivation of SUSY transformations for any {\it general} $\mathcal {N} = 2$ SUSY QM system.

One of the key differences between the (anti-)BRST and SUSY symmetry transformations is that, the (anti-)BRST symmetries are nilpotent 
as well as absolutely anticommuting in nature, whereas SUSY transformations are {\it only} nilpotent and the anticommutator of 
fermionic transformations produces an {\it additional} symmetry transformation of the theory. Due to this basic reason, we are theoretically 
forced to use the (anti-)chiral supervariables generalized on the (1, 1)-dimensional super-submanifolds of the
full (1, 2)-dimensional supermanifold. The latter is parametrized by the superspace coordinate
$Z^M = (t, \theta,\bar\theta)$ where $\theta,\,\bar\theta$ are the Grassmannian variables and $t$ is the time-evolution parameter .

The contents of the present investigation are organized as follows. In Sec. 2, we discuss about the symmetry transformations
associated with the specific $\mathcal{N} = 2$ SUSY QM model. It is to be noted that, there are {\it three} 
continuous symmetries associated with this particular model, in which {\it two} of them are fermionic and {\it one} is bosonic in nature.  
Sec. 3 is devoted for the derivation of one of the fermionic transformations by using the anti-chiral suparvariable approach. We derive the 
second SUSY fermionic symmetry by exploiting the chiral supervariable approach in Sec 4.  
In Sec 5, the Lagrangian of the  model is presented in terms of the (anti-)chiral supervariables and the geometrical interpretation for  
invariance of Lagrangian in terms of the Grassmannian derivatives ($\partial_\theta$ and $\partial_{\bar\theta}$) is explicated. 
Furthermore, we also represent the charges corresponding to the continuous symmetry transformations in terms of (anti-)chiral supervariables. 
In Sec 6, we show that the fermionic SUSY symmetry transformations satisfy the $\mathcal {N} = 2$ SUSY 
algebra, which is identical to the Hodge algebra obeyed by the cohomological operators of differential geometry. Thus, we show that this 
particular $\mathcal{N} = 2$  SUSY QM model is an example of Hodge theory. Finally, we conclude in Sec. 7, with remarks.

\section {Preliminaries: A specific $\mathcal {N} = 2$ SUSY QM model}

We begin with the action of a specific (0 + 1)-dimensional $\mathcal {N} = 2$ QM model [21]
\begin{eqnarray}
S = \int dt L_0 = \int dt\Big[  \Big(\frac{d\phi}{dt} + s \frac{\partial V}{\partial \phi} \Big )^2 
- i\, \bar\psi \Big( \frac{d}{dt} + s \frac{\partial^2 V}{\partial \phi \partial \phi} \Big) \psi \Big],
\end{eqnarray}where the bosonic variable $\phi$  and fermionic variables $\psi, \bar\psi$ are the functions of time-evolution parameter $t$,
 $V(\phi)$ is a general potential function and $s$ is an independent constant parameter. For algebraic convenience,
 we linearize the first term in (1) by introducing an auxiliary variable $A$. As a consequence, the 
action can be written as\footnote{ Henceforth, we denote  $\dot \phi =\frac{d \phi}{dt}, V' = \frac{\partial V}{\partial\phi},$ $V'' = \frac{\partial^2 V}{\partial\phi \partial\phi}$ in the text.}
\begin{eqnarray}
S = \int dt L = \int dt\Big[ i \Big(\dot\phi + s V' \Big ) A 
+  \frac{A^2}{2} - i\, \bar\psi \Big( \dot \psi + s\, \psi\, V''\Big) \Big],
\end{eqnarray}
where $A = -\,i\,(\dot\phi + s V')$ is the equation of motion. Using this expression for $A$ in (2), one can recover  the original action.

For the present QM system, we have the following off-shell nilpotent ($s^2_1 = 0, s^2_2 = 0$) SUSY 
transformations\footnote{We point out that the SUSY transformations in (3) are differ by an overall $i$-factor from [20].}: 
\begin{eqnarray}
&& s_1\, \phi = i\,\psi, \qquad \quad s_1 \,\psi = 0, \qquad  \quad s_1\, \bar \psi = i\,A, \qquad \quad s_1 \,A = 0, \nonumber\\
&& s_2\, \phi = i\,\bar\psi, \qquad s_2\, \bar\psi = 0, \qquad  s_2 \, \psi = i\,A - 2 \,s\, V', 
\qquad s_2\, A = 2 \,s\,\bar\psi\, V''. 
\end{eqnarray}
Under the above symmetry transformations, the Lagrangian in (2) transforms as 
\begin{eqnarray}
s_1 L =  0, \qquad\qquad
s_2 L = \frac{d}{dt}(- A\, \bar\psi ).
\end{eqnarray}
Thus, the action integral remains invariant (i.e., $s_1\, S = 0,\; s_2\, S = 0$). According to  Noether's theorem, the
continuous SUSY symmetry transformations $s_1$ and $s_2$ lead to the following conserved charges, respectively: 
\begin{eqnarray}
Q = -\psi\, A,\qquad \quad \bar Q = -\bar\psi \Big[ A + 2\,i\,s\, V' \Big],
\end{eqnarray}
The conservation of the SUSY charges (i.e., $\dot Q = 0,\, \dot{\bar Q} = 0$) can be proven by exploiting the  following 
Euler-Lagrange equations of motion:
\begin{eqnarray}
&& \dot A = s\, A\, V'' - s\, \bar\psi\, \psi\, V''',\qquad \dot{\bar\psi} = - s\,\bar \psi  V'',\nonumber\\
&&  A = -i\, (\dot\phi + s\,V'),\qquad \quad \dot{\psi} = - s \,\psi\, V''. 
\end{eqnarray}
It turns out that these charges are the generators of SUSY transformations (3). As one can explicitly check that the following relations are true
\begin{eqnarray}
s_1 \Phi =  i [\Phi, Q]_{\pm}, \qquad \; s_2 \Phi =  i [\Phi, \bar Q]_{\pm}, \qquad \quad \Phi = \phi, A, \psi, \bar\psi, 
\end{eqnarray} 
where the subscripts ($\pm$), on the square brackets, deal with the (anti)commutator depending 
on the variables being (fermionic)bosonic in nature.

It is to be noted that the anticommutator of the fermionic SUSY transformations ($s_1$ and $s_2$) 
lead to a bosonic symmetry ($s_\omega$), 
\begin{eqnarray}
&& s_\omega\, \phi = - 2( A + i\, s\, V'),\qquad \quad s_\omega\, \psi = -2\,i \,s\,\psi\,V'', \nonumber\\
&& s_\omega\,\bar\psi = 2\,i\,s\, \bar\psi\, V'', \qquad \quad  s_\omega\, A = 2\,i\,s\, (A\, V'' - \bar\psi\,\psi V'''). 
\end{eqnarray}
The application of  bosonic symmetry ($s_\omega$) on Lagrangian produces total time derivative,
\begin{eqnarray}
s_\omega L = (s_1\,s_2 + s_2\, s_1) L = \frac{d }{dt} \Bigl(-\,i \,A^2 \Bigr).
\end{eqnarray}   
Thus, according to the Noether's theorem, the above continuous bosonic symmetry leads to a bosonic conserved charge ($Q_\omega$) as follows,
\begin{eqnarray}
Q_\omega &=& -i\, A^2 + 2\,s\,A V' + 2\,s\,\psi\,\bar\psi\,V''\nonumber\\ 
&=& 2\,i \Bigl[ \frac{\Pi_{\phi}\,\Pi_{\phi}}{2} - s\,\Pi_{\phi}\, V' 
- s\,\psi\, \Pi_\psi\, V''\Bigr] \equiv(2\,i) H,
\end{eqnarray} 
where $\Pi_{\phi} = i\, A,\, \Pi_{\psi} = i\, \bar\psi$ are the canonical momenta corresponding to the variables $\phi,\, \psi$, respectively. 
It is clear that the bosonic charge $Q_\omega$ is the Hamiltonian $H$ (modulo a constant $2i$-factor) of our present model.

One of the important features of SUSY transformations 
is that the application of this bosonic symmetry must produce the time translation
of the variable (modulo a constant $2\,i$ factor), which can be checked as,
\begin{eqnarray}
s_\omega \Phi = \{s_1, s_2 \} \Phi = (2\,i) \dot \Phi, \qquad \quad \Phi = \phi, A, \psi, \bar\psi,
\end{eqnarray}
where in order to prove the sanctity of this equation, we have used the equations of motion mentioned in (6).

\section{Off-shell nilpotent SUSY transformations: Anti-chiral supervariable approach}

In order to derive the continuous transformation $s_1$,
we shall focus on the (1, 1)-dimensional super-submanifold (of general (1, 2)-dimensional supermanifold)  parameterized 
by the supervariable ($t, \bar\theta$). For this purpose, we impose supersymmetric invariant restrictions (SUSYIRs) on the anti-chiral supervariables.
We then generalize the basic  (explicit $t$ dependent) variables to their  anti-chiral supervariable counterparts:
\begin{eqnarray}
&& \phi (t) \longrightarrow {\tilde \Phi}(t, \theta, \bar\theta)|_{\theta = 0} 
\equiv  {\tilde \Phi}(t, \bar\theta) = \phi(t) + \bar\theta\, f_1(t), \nonumber\\
&& \psi (t) \longrightarrow {\tilde \Psi}(t, \theta, \bar\theta)|_{\theta = 0} 
\equiv {\tilde \Psi}(t, \bar\theta) = \psi(t) + i\,\bar\theta\, b_1(t), \nonumber\\
&&\bar\psi (t) \longrightarrow {\tilde {\bar\Psi}}(t, \theta,\bar\theta)|_{\theta = 0} 
\equiv {\tilde {\bar\Psi}}(t, \bar\theta) = \psi(t) + i\,\bar\theta\, { b}_2(t), \nonumber\\
&& A(t) \longrightarrow \tilde A (t, \theta,\bar\theta)|_{\theta = 0} 
\equiv \tilde A (t, \bar\theta) = A(t) + \bar\theta P (t),
\end{eqnarray}
where $f_1(t), P(t)$ and $ b_1(t),  b_2(t)$ are the fermionic and bosonic secondary variables, respectively.

It is observed from (3) that $s_1\, (\psi, A) = 0$ (i.e.,  both $\psi$ and $A$  are invariant under $s_1$). 
Therefore, we demand that the both variables should remain unchanged due to the presence of Grassmannian variable $\bar\theta$. 
As a result of the above restrictions, obtains,
\begin{eqnarray}
{\tilde\Psi} (t, \theta, \bar\theta)|_{\theta = 0} \equiv {\tilde\Psi} (t, \bar\theta) = \psi(t) \Longrightarrow  b_1 = 0,\nonumber\\
\tilde A (t, \theta, \bar\theta)|_{\theta = 0} \equiv  \tilde A (t, \bar\theta) = A(t)\Longrightarrow  P = 0.
\end{eqnarray}
We further point out that $s_1\, (\phi\, \psi) = 0$ and $s_1\, ({\dot \phi}\, {\dot \psi}) = 0$
 due to the fermionic nature of $\psi$ (i.e., $\psi^2 = 0$). Thus, these restrictions yield,
\begin{eqnarray}
&&  {\tilde \Phi}(t, \bar\theta)\, {\tilde \Psi} (t, \bar\theta) = \phi(t)\,\psi(t) \Longrightarrow f_1\, \psi = 0, \nonumber\\
&& {\dot {\tilde\Phi}}(t, \bar\theta)\, {\dot {\tilde \Psi}} (t, \bar\theta) 
= {\dot \phi}(t)\,\dot \psi(t) \Longrightarrow {\dot f}_1\, \dot\psi = 0. 
\end{eqnarray} 
The trivial solution for the above relationships is  $f_1 \propto \psi$, for algebraic convenience, 
we choose $ f_1 = i\, \psi$.   Here the $i$-factor has been taken due to the 
convention we have adopted for the present SUSY QM theory. Substituting the values of the secondary variables in the 
expansions of anti-chiral supervariables (12), one obtains,
\begin{eqnarray}
&& {\tilde \Phi}^{(ac)}(t, \bar\theta) = \phi(t) + \theta\,  (i \,\psi ) \equiv \phi(t) + \bar\theta \,\Big(s_1\, \phi(t)\Big),\nonumber\\
&& {\tilde \Psi}^{(ac)}(t, \bar\theta) = \psi(t) + \bar\theta\, (0) \equiv \psi (t) + \bar\theta\,\Big(s_1\, \psi(t) \Big) \nonumber\\
&& \tilde A^{(ac)} (t, \bar\theta) = A(t) + \theta (0) \equiv A(t) + \bar\theta\, \Big(s_1\, A(t)\Big).
\end{eqnarray}
The superscript $(ac)$ in the above represents the anti-chiral supervariables, obtained after the application 
of SUSIRs. Furthermore, we note that the 1D potential function $V(\phi)$ can be generalized to 
$\tilde V({\tilde \Phi}^{(ac)})$ onto  (1, 1)-dimensional super-submanifold as:
\begin{eqnarray}
V(\phi) \longrightarrow {\tilde V}({\tilde \Phi}^{(ac)}) = {\tilde V}^{(ac)}\bigl(\phi + \bar\theta (i\,\psi)\bigr) = V(\phi) 
+ \bar\theta\, \Big(i\,\psi\, V' \Big) \equiv V (\phi) + \bar\theta\, \Big(s_1\,V(\phi) \Big),
\end{eqnarray}
where we have used the expression of anti-chiral supervariable ${\tilde \Phi}^{(ac)} (t, \bar\theta)$ as given in (15).

In order to find out the SUSY transformation for $\bar\psi$, it can be checked that the application of $s_1$ on the 
following vanishes:  
\begin{eqnarray}
s_1\,\big[ i \,(\dot\phi + s\,  V')\, A - i\, \bar\psi \,( \dot \psi + s\, \psi V'')  \big] = 0.
\end{eqnarray}
As a consequence, the above can be used as a SUSYIR and we replace the ordinary variables by their anti-chiral supervariables as:
\begin{eqnarray}
 \big[ i\big({\dot {\tilde \Phi}^{(ac)}} + s\, \tilde V'^{(ac)}\big ) {\tilde A}^{(ac)} 
 - i {\tilde{\bar\Psi}} \big( {\dot{\tilde\Psi}^{(ac)} } + s {\tilde\Psi}^{(ac)} {\tilde V''^{(ac)}} \big)   \big ]
 = \big[ i (\dot\phi + s  V') A - i \bar\psi\, ( \dot \psi + s \psi V'')  \big].
 \end{eqnarray}
After doing some trivial computations, we obtain $ b_2 = A $.
Recollecting all the value of  secondary variables and substituting them into (12), we finally yield the 
following anti-chiral supervariable expansions
\begin{eqnarray}
&& {\tilde \Phi}^{(ac)}(t, \bar\theta) = \phi(t) + \bar\theta\,  (i \psi) \equiv \phi(t) + \bar\theta\,\Big(s_1\, \phi(t)\Big),\nonumber\\
&& {\tilde \Psi}^{(ac)}(t, \bar\theta) = \psi(t) + \bar\theta\, (0) \equiv \psi (t) + \bar\theta\,\Big(s_1\, \psi(t)\Big), \nonumber\\
&& {\tilde {\bar\Psi}} ^{(ac)}(t, \bar\theta) = \bar\psi (t) + \bar\theta \,(i A) \equiv \bar\psi (t) + \bar\theta\,\Big(s_1\, \bar\psi(t)\Big), \nonumber\\
&& {\tilde A} ^{(ac)}(t, \bar\theta) = A(t) + \bar\theta\, (0) \equiv A(t) + \bar\theta\,\Big(s_1 \,A(t)\Big).
\end{eqnarray}
Finally, we have derived explicitly, the SUSY transformation $s_1$ for all the 
variables by exploiting SUSY invariant restrictions on the chiral supervariables. These symmetry transformations are:
\begin{eqnarray}
&& s_1\, \phi =  i\,\psi,\quad  \; s_1\, \psi = 0,\quad\; s_1\,\bar \psi = i\,A, \nonumber\\
&& s_1\, A  = 0, \qquad \quad s_1\, V = i\, \psi\, V'.
\end{eqnarray}
It is worthwhile to mention here that for the anti-chiral supervariable expansions given in (11), we have the following 
relationship between the Grassmannian derivative $\partial_{\bar\theta}$ and SUSY transformations $s_1$:
\begin{eqnarray}
\frac{\partial}{\partial \bar\theta}\, {\tilde \Omega}^{(ac)} (t,\theta, \bar\theta)|_{\theta = 0} 
\equiv \frac{\partial}{\partial \bar\theta}\, {\tilde\Omega}^{(ac)} (t,\bar\theta) = s_1\, \Omega (t), 
\end{eqnarray}
where  $\Omega^{(ac)} (t,\bar\theta)$ is the generic supervariable obtained by exploiting the 
SUSY invariant restriction on the anti-chiral supervariables.
It is easy to check from the above equation that the symmetry transformation ($s_1$) for any generic variable  $\Omega(t)$ is equal to the
translation along the $\bar\theta$-direction of the anti-chiral supervariable.
Furthermore, it can also be checked that nilpotency of the Grassmannian derivative $\partial_{\bar\theta}$ 
(i.e. $\partial^2_{\bar\theta} = 0$) implies $s^2_1 = 0$.

\section{Off-shell nilpotent SUSY transformations: Chiral supervariable approach}

For the derivation of second fermionic SUSY transformation $s_2$, we concentrate on the 
chiral super-submanifold parametrized by the supervariables $(t, \theta)$.
Now all the ordinary variables (depending explicitly on $t$) are generalized to a (1, 1)-dimensional chiral 
super-submanifold as:
\begin{eqnarray}
&& \phi(t) \longrightarrow {\tilde \Phi}(t,\theta, \bar\theta)|_{\bar\theta = 0} \equiv {\tilde \Phi}(t, \theta) = \phi(t) + \theta\,  {\bar f}_1(t), \nonumber\\
&& \psi(t) \longrightarrow  {\tilde \Psi}(t, \theta, \bar\theta)|_{\bar\theta = 0} \equiv {\tilde \Psi}(t, \theta) 
= \psi(t) +  i\,\theta\,  {\bar b}_1(t), \nonumber\\
&& \bar\psi(t) \longrightarrow {\tilde {\bar \Psi}}(t, \theta, \bar\theta)|_{\bar\theta = 0} \equiv {\tilde {\bar \Psi}}(t, \theta) = \psi(t) + i\,\theta\, {\bar b}_2(t), \nonumber\\
&& A(t) \longrightarrow  {\tilde A}(t, \theta, \bar\theta)|_{\bar\theta = 0} \equiv A(t,\theta) = A(t) + \theta\, {\bar P}(t).
\end{eqnarray}
In the above, secondary variables ${\bar f }_1(t),\, {\bar P}(t)$ and ${\bar b}_1(t),\, {\bar b}_2(t)$ are fermionic and bosonic variables, respectively. 
We can derive the values of these secondary variables in terms of the basic variables, by exploiting the
power and potential of SUSY invariant restrictions.

It is to be noted from (3) that  $\bar\psi$ does not transform under SUSY 
transformations $s_2$ (i.e. $s_2\, \bar\psi = 0$) so the variable $\bar\psi$ would remain unaffected by the presence of Grassmannian variable $\theta$.
As a consequence, we have the following:
\begin{eqnarray}
{\tilde{\bar\Psi}} (t, \theta, \bar\theta)|_{\bar\theta = 0} \equiv {\tilde{\bar\Psi}} (t, \theta) = \bar \psi(t) \Longrightarrow  {\bar b}_2 = 0.
\end{eqnarray}
Furthermore, we observe that $s_2\, (\phi\, \bar \psi) = 0$ and $s_2\, (\dot {\phi}\, \dot {\bar\psi}) = 0$ 
due to the fermionic nature of $\bar\psi$. Generalizing these invariant restrictions to the chiral super-submanifold, we have the following SUSYIRs in the following forms, namely;
\begin{eqnarray}
{\tilde \Phi}(t,\theta)\, {\tilde{\bar\Psi}} (t, \theta) = \phi(t)\,\bar\psi(t),\nonumber\\
\dot {\tilde{\bar \Phi}}(t, \theta)\, \dot {\bar\Psi} (t, \theta) = {\dot \phi}(t)\,\dot {\bar\psi}(t).   
\end{eqnarray} 
After putting the expansions for the supervariable (22) in the above, we get 
\begin{eqnarray}
{\bar f}_1\, \bar\psi = 0, \qquad \quad {\dot {\bar f}}_1\, \dot{\bar\psi} = 0.
\end{eqnarray} 
The  solution for the above relationship is ${\bar f}_1 = i\,\bar\psi$. 
Substituting the value of secondary variables in the chiral supervariable expansions (21), we obtain the following expressions:
\begin{eqnarray}
&& {\tilde \Phi}^{(c)}(t, \theta) = \phi(t) + \theta\, (i\,\bar\psi) \equiv \phi(t) + \theta\, \Big(s_2\, \phi(t)\Big),\nonumber\\
&& {\tilde{\bar \Psi}}^{(c)}(t, \theta) = \bar\psi(t) + i \theta\, (0) \equiv  \psi (t) + \theta\,\Big(s_2\, \bar\psi (t)\Big),
\end{eqnarray}
where superscript $(c)$ represents the chiral supervariables obtained after the application of SUSYIRs.
Using (26), one can generalize  $V(\phi)$  to ${\tilde V}({\tilde \Phi}^{(c)})$  onto 
the (1, 1)-dimensional chiral super-submanifold as
\begin{eqnarray}
V(\phi) \longrightarrow {\tilde V}(\tilde\Phi^{(c)}) = {\tilde V}^{(c)}(\phi + \theta (i\,\bar\psi)) = V(\phi) + \theta\, \Big(i\,\bar\psi\,V'(\phi)\Big)
\equiv V (\phi) + s_2\,\Big(V(\phi)\Big),
\end{eqnarray}
where we have used the expression of  chiral supervariable ${\tilde\Phi}^{(c)} (t, \theta)$ given in (26).

We note that $s_2\,[i\,A - 2\,s\, \bar\psi\, V'] = 0$ because of the nilpotency of $s_2$  [cf. (3)]. 
Thus, we have the following SUSYIR in our present theory:
\begin{eqnarray}
i\,\tilde A^{(c)} - 2\,s\,{\tilde{\bar\Psi}} \, {\tilde V}'^{(c)} = [i\,A - 2\,s\, \bar\psi\, V'].
\end{eqnarray}
This restriction serves our purpose for the  derivation of SUSY transformation of $A(t)$. Exploiting the above restriction, 
we get the value of secondary variable ${\bar P}(t)$ in terms of basic 
variables as: $\bar P = 2\, s\,\bar\psi\, V''$.

It is important to note that the following sum of the composite variables are invariant under $s_2$, namely;
\begin{eqnarray}
s_2 \big[i\, s\,  V' \,A + \frac{1}{2} A^2 - i\,s\, \bar\psi\,\psi\,V'' \big] = 0.
\end{eqnarray}
In order to calculate the fermionic symmetry transformation corresponding to variable $\psi$, we use the 
above relationship as a SUSYIR:
\begin{eqnarray}
 i\,s{\tilde V}'^{(c)} \,{\tilde A}^{(c)} + \frac{1}{2}{\tilde A}^{2(c)}
- i\,s\,{\tilde  {\bar\Psi}^{(c)}} \, {\tilde\Psi}\,{\tilde V}''^{(c)} = i\, s\,  V' \,A + \frac{1}{2} A^2 
- i\,s\, \bar\psi\,\psi\,V'',
\end{eqnarray}
after some computations, one gets $ {\bar b}_1 =  A + 2\,i\,s\, V'$.

Recollecting all the values of secondary variables and substituting them into (21), we have the 
expansions of the anti-chiral supervariables as:  
\begin{eqnarray}
&& {\tilde \Phi}^{(c)}(t, \theta) = \phi(t) + \theta\, \Big(i\,\bar\psi \Big) \equiv \phi(t) + \theta\,\Big(s_2\, \phi(t) \Big),\nonumber\\
&& {\tilde \Psi}^{(c)}(t, \theta) = \psi (t) + \theta\, \Big( i\,A - 2\,s\, V'\Big) 
\equiv \psi (t) + \theta\,\Big(s_2 \psi (t) \Big), \nonumber\\
&& {\tilde{\bar \Psi}}^{(c)}(t, \theta) = \bar\psi(t) + i\, \theta\, \Big(0 \Big) \equiv  \psi (t) + \theta\, \Big(s_2\,\bar\psi (t) \Big),\nonumber\\
&& {\tilde A}^{(c)}(t, \bar\theta) = A(t) + \theta \,\Big( 2\, s\,\bar\psi\, V'' \Big) 
\equiv  A(t) + \theta\, \Big(s_2  A(t) \Big).
\end{eqnarray}
Finally, the supersymmetric transformations  ($s_2$) for all the basic and auxiliary variables are listed as,
\begin{eqnarray}
&& s_2\, \phi =  i\,\bar\psi,\quad \;  s_2\, \bar\psi = 0, \quad \; s_2\, \psi =  i\,A - 2\,s\, V', \nonumber\\
&& s_2\, V = i\, \bar\psi\, V', \qquad \quad s_2\, A = 2\, s\,\bar\psi\, V''.
\end{eqnarray}
It is important to point here that, we have the following mapping between the Grassmannian derivative 
($\partial_{\bar\theta}$) and the symmetry transformation $s_2$:
\begin{eqnarray}
\frac{\partial}{\partial\theta}\, {\tilde \Omega}^{(c)} (t,\theta, \bar\theta)|_{\bar\theta = 0} 
= \frac{\partial}{\partial\theta}\, {\tilde\Omega}^{(c)} (t,\theta) = s_2\, \Omega (t),
\end{eqnarray}
where ${\tilde \Omega}^{(c)} (t,\theta)$ is the generic chiral supervariables obtained after the application of
supersymmetric invariant restrictions and $\Omega (t)$ denotes the basic variables of our present QM theory.
The above equation captures the geometrical interpretation of transformation ($s_2$), in terms of the Grassmannian
derivative ($\partial_\theta$) because of the fact that, the translation along $\theta$-direction of chiral supervariable is 
equivalent to the symmetry transformation ($s_2$) of the same basic variable.  
We observe from (33) that the nilpotency of SUSY transformation $s_2$ (i.e., $s^2_2 = 0$) can be generalized in terms of 
Grassmannian derivative $\partial^2_{\theta} = 0$.

\section{Invariance and off-shell nilpotency: Supervariable approach }

It is interesting to note that by exploiting the expansions of the supervariables (11), the Lagrangian in (2) can be expressed  in terms 
of the anti-chiral supervariables as 
\begin{eqnarray}
L \Longrightarrow L^{(ac)} = i\Big( \dot{\tilde \Phi}^{(ac)} + s \,{\tilde V}^{'(ac)} \Big) {\tilde A}^{(ac)} 
+ \frac{1}{2}{\tilde A}^{(ac)}\,{\tilde A}^{(ac)} - i\, {\tilde {\bar\Psi}^{(ac)}} \Big( \dot{\tilde \Psi}^{(ac)} 
+ s\, {\tilde \Psi}^{(ac)} {\tilde V}^{''(ac)}\Big),
\end{eqnarray}
In the earlier section, we have shown that SUSY transformation ($s_1$) and translational generator ($\partial_{\bar\theta}$) 
are geometrically related to each-other (i.e., $s_1\leftrightarrow \partial_{\bar\theta}$). As a consequence, 
one can also capture the invariance of the Lagrangian in the following fashion: 
\begin{eqnarray}
 \frac{\partial}{\partial\bar\theta} L^{(ac)} = 0 \qquad  \Longleftrightarrow \qquad s_1\, L  = 0.
\end{eqnarray}
Similarly, the Lagrangian (2) can also be  written in terms of the chiral supervariables as: 
\begin{eqnarray}
L \Longrightarrow L^{(c)} = i\Big(\dot{\tilde \Phi}^{(c)} + s {\tilde V}^{'(c)} \Big) {\tilde A}^{(c)} + \frac{1}{2}{\tilde A}^{(c)}\,{\tilde A}^{(c)}
- i\, {\tilde {\bar\Psi}^{(c)}} \Big( \dot{\tilde {\Psi}}^{(c)} + s\, {\tilde {\Psi}}^{(c)} {\tilde V}^{''(c)}\Big).
\end{eqnarray}
Since the fermionic symmetry ($s_2$) is geometrically connected with the translational generator ($\partial_{\theta}$), 
therefore, the invariance of the Lagrangian can also be geometrically interpreted as follows:
\begin{eqnarray}
 \frac{\partial}{\partial\theta} L^{(c)} = \frac{d}{dt} \Big[-\psi\, A \Big] \qquad \Longleftrightarrow \qquad s_2\, L  = \frac{d}{dt}\Big[-\psi\, A \Big].
\end{eqnarray}
As a result, the action integral $S = \int dt \, L^{(c)}|_{\theta = 0}$ remains invariant.

We point out that the conserved charges $Q$ and $\bar Q$ corresponding to the continuous symmetry transformations 
$s_1$ and $s_2$ can also be expressed as,
\begin{eqnarray}
&& Q = s_1 \,[-i\, \psi\, \bar\psi] = - \psi \, A \qquad \Longleftrightarrow \qquad Q =  \frac{\partial}{\partial {\bar\theta}} 
\Big[ -i\, {\tilde \Psi}^{(ac)}\, {\tilde {\bar\Psi}^{(ac)}} \Big], \nonumber\\
&& \bar Q = s_2\, [i \,\psi\, \bar\psi] = - \bar\psi \big[ A + i \,s\,  V' \big] \qquad \Longleftrightarrow \qquad \bar Q = \frac{\partial}{\partial \theta} 
 \Big[ i\, {\tilde \Psi}^{(c)}\, {\tilde {\bar\Psi}^{(c)}} \Big].
\end{eqnarray} 
The nilpotency properties of the above charges can be shown in a straightforward manner with the help of symmetry properties,
\begin{eqnarray}
&& s_1\, Q = +i\,\{ Q, Q \} = 0 \quad \Longrightarrow \quad  Q^2 = 0,  \nonumber\\
&& s_2 \,\bar Q = +i\,\{\bar Q, \bar Q \} = 0 \quad \Longrightarrow \quad  {\bar Q}^2 = 0.
\end{eqnarray}
In the language of translational generators, these properties can be written as  $\partial_{\bar \theta}\, Q = 0 \Rightarrow \bar Q^2 = 0$ and 
$\partial_{\theta}\, \bar Q = 0 \Rightarrow Q^2 = 0$. These relations hold due to the nilpotency of the Grassmannian derivatives 
(i.e., $\partial^2_{\bar \theta} = 0, \partial^2_\theta = 0$).

\section{$\mathcal {N} = 2$ SUSY algebra and its interpretation}

We observe that under the following discrete symmetry, 
\begin{eqnarray}
&& t \rightarrow t, \qquad \quad \phi \rightarrow \phi, \qquad \quad \psi\rightarrow \bar\psi, \nonumber\\
&&  \bar\psi \rightarrow \psi, \qquad \; s \rightarrow -\,s, \qquad\; A \rightarrow A + 2\,i\,s\, V',
\end{eqnarray}
the Lagrangian ($L$) transforms as: $L \longrightarrow L + \frac{d}{dt}[i\, \bar\psi\, \psi -\,2\,s \,V]$. 
Hence, action integral (2) of the SUSY QM system remains invariant.
It is to be noted that, the above discrete symmetry transformations are important because they relate the two SUSY transformations ($s_1, s_2$):
\begin{eqnarray}
s_2 \,\Omega = \pm *\, s_1 \, * \Omega, 
\end{eqnarray}
where $\Omega$ is the generic variables present in the model.
It is to be noted that, generally, the ($\pm$) signs are governed by the two successive operations of the discrete symmetry on the variables as: 
\begin{eqnarray}
* \,(*\, \Omega) = \pm \Omega.
\end{eqnarray}
In the present case, {\it only} the ($+$) sign  will occur for all the variables (i.e. $\Omega = \phi, \psi, \bar\psi, A$). 
It can be easily seen that the relationship (41) is analogous to the relationship, $\delta = \pm \, *\,d\,*$ of 
differential geometry (where $d$ and $\delta$ are the exterior and co-exterior derivative, respectively and ($*$) is the Hodge 
duality operation).

We now focus on the physical identifications of the de Rham cohomological operators of 
differential geometry in terms of the symmetry transformations. It can be explicitly checked 
that the continuous symmetry transformations, together with discrete symmetry for our SUSY QM model, satisfy 
the following algebra [17-20, 22-24]:
\begin{eqnarray}
&& s^2_1 = 0, \qquad \; s^2_2 = 0, \qquad \; \{s_1, \, s_2 \} =  (s_1 + s_2 )^2 = s_\omega, \nonumber\\
&& [s_1, \, s_\omega ] = 0, \qquad \quad  [s_2,\, s_\omega ] = 0, \qquad s_2 = \pm *s_1*,
\end{eqnarray}  
identical to the algebra obeyed by the de Rham cohomological operators ($d, \delta, \Delta$) [25-29],
\begin{eqnarray}
&&d^2 = 0, \qquad \quad \delta^2 = 0, \qquad \{d, \delta \} = ( d + \delta )^2 = \Delta, \nonumber\\
&& [d, \Delta ] = 0, \quad \quad [\delta, \Delta ] = 0, \qquad \delta = \pm \,\star \, d\,\star.
\end{eqnarray}
Here $\Delta$ is the Laplacian operator. 
From the above equations (43) and (44), we can identify the exterior derivative with $s_1$ and co-exterior derivative $\delta$ with $s_2$. 
The discrete symmetry (40) provides the {\it analogue} of Hodge duality ($\star$) operation of 
differential geometry. In fact, there is a one-to-one mapping between the symmetry transformations and the de Rham cohomological operators. 
It is also clear from (43) and (44) that, the bosonic symmetry ($s_\omega$) 
and Laplacian operator ($\Delta$) are the Casimir operators of the algebra given in (43) and (44), respectively.
Thus, our present ${\cal N} = 2$ SUSY model provides a model for Hodge theory. 
Furthermore,  a similar  algebra given in (44) is also 
satisfied by the conserved charges $Q, \bar Q$ and $Q_\omega$: 
\begin{eqnarray}
&& Q^2 =  0, \quad\; {\bar Q}^2 = 0, \quad\; \{Q, \bar Q \} = Q_\omega = 2i\,H, \nonumber\\
&& [Q, H] = 0, \qquad \quad [\bar Q, H] = 0.
\end{eqnarray}
In the above, we have used the canonical quantum (anti)commutation relations $[\phi, A ] = 1$ and $\{\psi,\, \bar\psi \} = 1$.
It is important to mention here that, the bosonic charge (i.e., the Hamiltonian of the theory modulo a $2i$-factor) is the Casimir operator
in the algebra (46).

Some crucial properties related to the de Rham cohomological operators ($d,\, \delta,\, \Delta$) can be captured
by these charges. For instance, we observe from (45) that the Hamiltonian is the Casimir operator of the algebra. Thus, it can be easily seen
that $H\,Q = Q\, H$ implies that $Q\, H^{-1} = H ^{-1}\, Q$, if the inverse of the Hamiltonian exists. Since we are dealing with the no-singular Hamiltonian, we presume that the Casimir operator has it well-defined inverse value. By exploiting (45) it can be seen 
\begin{eqnarray}
&& \Big[ \frac{Q\, \bar Q}{H}, Q \Big] = Q, \qquad \quad \Big[ \frac{Q\, \bar Q}{H}, \bar Q \Big] = - \bar Q, \nonumber\\
&& \Big[ \frac{\bar Q\, Q}{H}, Q \Big] = - Q, \qquad \quad \Big[ \frac{\bar Q\, Q}{H}, \bar Q \Big] =  \bar Q.
\end{eqnarray}
Let us define an eigenvalue equation $\frac{Q\, \bar Q}{H} |\chi \rangle_p = p |\chi \rangle_p$ where $|\chi \rangle_p$ is the
quantum Hilbert state with eigenvalue $p$. By using the algebra (46) one can verify the following:
\begin{eqnarray}
&&\Big(\frac{Q\, \bar Q}{H} \Big)\, Q |\chi \rangle_p = (p + 1)\, |\chi \rangle_p, \nonumber\\
&&\Big(\frac{Q\, \bar Q}{H} \Big)\, \bar Q |\chi \rangle_p = (p - 1)\, |\chi \rangle_p, \nonumber\\
&&\Big(\frac{Q\, \bar Q}{H} \Big)\, H |\chi \rangle_p = p \, |\chi \rangle_p, \nonumber\\
\end{eqnarray}
As a consequence of the equation (47), it is evident that $Q |\chi \rangle_p,\, \bar Q |\chi \rangle_p$ and $H |\chi \rangle_p$ has the 
eigenvalues $(p+1), \, (p-1)$ and $p$ respectively w.r.t to the operator $\frac{Q\, \bar Q}{H}$.

The above equation provides a connection between the conserved charges ($Q,\, \bar Q, \, H$) and de Rham cohomological 
operators ($d\, \delta, \, \Delta$) because, as we know that the action of $d$ on a given form increases the degree of the form by one, 
whereas application of $\delta$ decreases the degree by one unit and operator $\Delta$ keeps intact the degree of a form. These important properties can be
realized by the charges ($Q, \bar Q,\, Q_\omega$), where the eigenvalues and eigenfunctions play the key role [22].

\section{Conclusions}

In summary, exploiting the supervariable approach, we have derived the off-shell nilpotent 
symmetry transformations for the $\mathcal {N} = 2$ SUSY QM system. This has been explicated through the
1D SUSY invariant quantities, which remain unaffected due to the presence of the Grassmannian variables
$\theta$ and $\bar\theta$. Furthermore, we have provided the geometrical interpretation of the SUSY transformations
($s_1$ and $s_2$) in terms of the translational generators ($\partial_{\bar\theta}$ and $\partial_\theta$) along the Grassmannian
directions $\bar\theta$ and $\theta$, respectively. Further, we have expressed the
Lagrangian in terms of the (anti-)chiral supervariables and the invariance of the Lagrangian under continuous
transformations ($s_1, s_2$) has been shown within the translations generators along ($\theta, \bar\theta$)-directions.
The conserved SUSY charges corresponding to the fermionic symmetry transformations have been expressed in terms of 
(anti-)chiral supervariables and the Grassmannian derivatives. The nilpotency of fermionic charges have been captured 
geometrically, within the framework of supervariable approach by the Grassmannian derivatives.

Finally, we have shown that the algebra satisfied by the continuous symmetry transformations $s_1,\, s_2$ and $s_\omega$ 
(and corresponding charges) is exactly analogous to the Hodge algebra obeyed by the de Rham cohomological operators 
($d,\, \delta$ and $\Delta$) of differential geometry. The discrete symmetry of the theory provides physical 
realization of the Hodge duality ($*$) operation. Thus, the present $\mathcal {N} = 2$ SUSY QM model provides a model for Hodge theory.\\
\vskip 1.4cm

\noindent
{\bf Acknowledgements} \\

\noindent
AS would like to thank Prof. Prasanta K. Panigrahi for reading the manuscript
as well as for offering the valuable inputs on the topic of our present investigation. AS is also thankful to Dr. Rohit Kumar
for his important comments during the preparation of the manuscript.\\

\noindent
{\bf Conflict of Interest Statement}\\ 

\noindent
Author(s) declare(s) that there is no conflict of interest regarding the publication of this paper. 

\end{document}